# Above-gap Conductance Anomaly Studied in Superconductor-graphene-superconductor Josephson Junctions

Jae-Hyun Choi[1], Hu-Jong Lee[1,3,*], and Yong-Joo Doh[1,2,*]

[1]*Department of Physics, Pohang University of Science and Technology, Pohang 790-784*

[2]*Department of Display and Semiconductor Physics, Korea University Sejong Campus, Chungnam 339-700*

[3]*National Center for Nanomaterials Technology, Pohang 790-784*

**ABSTRACT**

We investigated the electrical transport properties of superconductor-graphene-superconductor (SGS) Josephson junctions. In low voltage bias, we observed conventional proximity-coupled Josephson effect, such as the supercurrent flow through the graphene, sub-gap structure of differential conductance due to Andreev reflection, and periodic modulation of the critical current $I_c$ with perpendicular magnetic field $H$ to the graphene. In high bias above the superconducting gap voltage, however, we also observed an anomalous jump of the differential conductance, the voltage position of which is sensitive to the backgate voltage $V_g$. Our extensive study with varying $V_g$, temperature, and $H$ reveals that the above-gap structure takes place at a characteristic power $P^*$, which is irrespective of $V_g$ for a given junction. Temperature and $H$ dependences of $P^*$ are well explained by the increase of the electron temperature in graphene.

# I. INTRODUCTION

Since first discovered in 2004 [1] graphene, a mono-atomic-layer honeycomb array of carbon atoms, has been intensively studied both theoretically and experimentally. Owing to its unique linear dispersion relation for low carrier energies, graphene leads to new physical phenomena associated with massless relativistic Dirac fermionic nature and chirality of carriers [2,3]. Moreover, a superconductor-graphene (SG) junction offers a unique system to investigate the interplay between the superconductivity and the relativistic quantum electrodynamics [4,5]. Unusual superconducting-proximity phenomena are theoretically predicted in the SG junctions [6,7]. To date, however, only a few groups [4,5,8] were successful to observe the superconducting proximity effect in superconductor-graphene-superconductor (SGS) junctions; *i.e.*, the supercurrent flow, multiple peaks of differential conductance ($dI/dV$), and magnetic-field-dependent modulation of the critical current (*i.e.*, the Fraunhofer pattern). Experimental difficulties are attributed to technical obstacles, such as forming a highly transparent contacts between graphene and superconducting electrodes, and establishing a noise-free measurement system, in particular, in the range of radio frequency or above.

Here, we report the fabrication of superconductor-graphene-superconductor (SGS) junctions and the successful observation of the superconducting proximity effect in graphene, such as an existence of the supercurrent, conductance enhancement due to multiple Andreev reflection (MAR), oscillating critical current $I_c$ with magnetic field ($H$). In addition, we observed the above-gap structure; an anomalous jump of $dI/dV$ occurring at high bias voltage ($V$) above the superconducting energy gap ($2\Delta/e$), where $e$ is an electric charge. The backgate voltage ($V_g$), temperature, and $H$ dependences of the above-gap structure indicate that the increase of the electron temperature $T_e$ in graphene due to Joule heating is responsible for the above-gap structure.

# II. EXPERIMENTS

For device fabrication, mono-layer graphene was mechanically exfoliated from natural graphite single crystals onto highly electron-doped Si substrate with 300-nm-thick thermally oxidized surface layer. Superconducting Al electrodes were formed on the graphene layer by conventional electron-beam lithography and subsequent electron-beam evaporation of a Ti/Al/Au multilayer (10/70/5 nm for the device D1; refer to Table I for other devices) under the base pressure of 2 x $10^{-7}$ Torr. Ti was used for better adhesion of Al to the substrate, while the Au capping layer was to protect Al from oxidation.

In measurements we minimize the contact resistance between the graphene and superconducting electrodes by delicately tuning the electron-beam lithographic parameters (for the dose of the writing beam and the developing time of the resist) and the deposition rate of the metallic electrodes. The radio-frequency noise was also reduced by adopting the multi-stage filtering scheme; two-stage RC filters in series with leads of a device and silver-powder filters in the cryogenic environment of $T < 1$ K in conjunction with π-filters arranged at room temperature. In addition, a small magnetic field was applied to cancel out any residual magnetic field in a cryostat for zero-field measurements.

Figure 1(a) shows an optical micrograph of the representative SGS junction device. For the device D1, the spacing ($L$) between the superconducting electrodes is ~300 nm while the width ($W$) is 3.2 μm (the boundary of graphene is denoted by a broken line). Dimensions of other devices are listed in Table I. A backgate voltage $V_g$ was applied to the highly electron-doped Si substrate to modulate the carrier density in graphene to be $n = \alpha |V_g|$, where $\alpha$~7.3 x $10^{10}$ cm$^{-2}$·V$^{-1}$ for a 300-nm-thick $SiO_2$ layer [1] on the surface of the substrate. Measurements were carried out with the base temperature at $T = 10$ mK, adopting a standard two-terminal configuration and the conventional ac lock-in technique [9]. In all our devices the contact resistance between the graphene and an electrode was less than 1 Ω, which was sufficiently smaller than the resistance of the graphene layer of an order of a few hundred Ω. Figure 1(b) shows the schematic cross-sectional view of the measurement configuration for an SGS junction.

## III. RESULTS AND DISCUSSION

Figure 1(c) shows the variation of the sheet resistance ($R_\square$) as a function of $V_g$ in the normal state at 0.5 K. $R_\square$ gradually increases along with the decrease of the carrier density as $V_g$ approaches the charge neutrality point or the Dirac point ($V_D$) [2-3]. But the value of $R_\square$ remains finite even for $V_g$ corresponding to the vanishing carrier density, *i.e.*, $V_g = V_D$, which is believed to be due to fluctuation of charge carriers by the presence of "puddles" [10]. The $V_g$ dependence of $R_\square$ renders the mobility to be $\mu \sim 658 - 1400$ cm$^2$/Vsec with the corresponding mean-free path of $l_{mean} \sim 16 - 40$ nm. For the estimation, we used the relation $\mu = \sigma/ne$ and $l_{mean} = m_e v_F \sigma/ne^2$, where $\sigma$, $n$, $m_e$, and $v_F$ are the sheet conductance, carrier density of graphene, electron mass, and Fermi velocity in graphene, respectively [1]. The single-layeredness of graphene is confirmed by the quantized conductance plateaus of $G = R_\square^{-1} = \nu e^2/h$ [2-3] in a perpendicular magnetic field of $H = 10$ T at temperature $T = 10$ mK [see the inset of Fig. 1(c)], where $\nu = 2, 6$ are filling factors. The dip-like feature for $\nu = 6$ is attributed to the low aspect ratio ($L/W$) of the device [11]. Four other samples tested in this study showed similar characteristics. In this report, main experimental results are presented in detail for the device D1. The superconducting transition temperature of Al electrodes of our devices was found to be $T_c \sim 0.38–0.84$ K (see Table I for details), which depended on the respective preparation condition of the Al film. The possible explanation for the $T_c$ reduction is discussed below.

Current-voltage (*I-V*) characteristics of the junction D1 for $T = 10$ mK in Fig. 2(a) clearly display the existence of the supercurrent, the maximum value of which varies with $V_g$, along with the hysteresis [see the inset of Fig. 2(a)]. In the main panel of Fig. 2(a), for simplicity, we plot the *I-V* characteristics only for uni-directional current-bias sweeping; sweeping up from -0.2 μA to +0.2 μA. The magnitude of the supercurrent ($I_c$) depends on the variation of the carrier density tuned by $V_g$. Figure 2(b) shows details of the $V_g$ dependence of the critical current $I_c$, which becomes maximally suppressed near the Dirac point ($V_D = -23$ V) but remains finite at ~10 nA. The incomplete vanishing of $I_c$ at $V_D$ is consistent with previous reports [4,5,8]. The magnitude of $I_c$ increases as $V_g$ moves away from $V_D$; $I_c$ reaches ~80 nA for $V_g = +30$ V. A Cooper pair is known to be transmitted into the graphene layer either as an electron pair or as a hole pair, depending on the value of $V_g$ relative to $V_D$ [4,5]; for $V_g > V_D$ an electron-like pair forms in the conduction band and for $V_g < V_D$ a hole-like pair forms in the valence band. It is also noted that the overlaid $G(V_g)$ curve looks like the $I_c(V_g)$ curve [see the black solid line in Fig. 2(b)]. This indicates that the $I_c$ variation with $V_g$ is mainly caused by the carrier density modulation in graphene with $V_g$ rather than by the transparency change at the SG interface.

To clearly confirm the occurrence of the dc Josephson effect in the SGS junction, an external-magnetic field $H$ was applied perpendicular to the graphene layer. We observed an oscillating behavior of $I_c$ with respect to $H$. The corresponding $H$ dependence of $I_c$ is displayed in Fig. 2(c), which reveals a periodic modulation of $I_c$ with $H$. Our experimental data are in good agreement with the theoretical expectation of $I_c(H) = I_c(0)\sin[2\pi(\Phi/\Phi_0)]/(\Phi/\Phi_0)$ [see the black broken line in Fig. 2(c)], where $\Phi$ is the magnetic flux threading the intermediate graphene layer and $\Phi_0 = h/2e$ the flux quantum in terms of the Planck's constant ($h$), with the field periodicity of $H^* = 0.61$ mT. In estimating $H^*$ one should take into account of the penetration of magnetic field into the Al superconducting electrodes by the amount of the London penetration depth $\lambda_L = 0.38$ μm as depicted by the solid line in the inset of Fig. 2(c).

As the bias current ($I$) exceeds $I_c$, an $I$-$V$ curve exhibits a voltage jump to a resistive state. The differential conductance ($dI/dV$) vs $V$ curve in this finite resistive state, taken by lock-in technique, exhibits multiple peaks as shown in Fig. 2(d). The overshooting $dI/dV$ near zero bias is due to the existence of supercurrent in the SGS junction, while the overall enhancement of $dI/dV$ overlaid with multiple peaks is caused by the Andreev reflections (AR) [12]. The AR process [13] occurs as an electron with energy lower than the gap value ($eV < \Delta$), incident from a normal-metallic side to a highly-transparent normal-metal–superconductor (NS) interface, is retroreflected as a phase-conjugated hole while a Cooper pair is formed and transmitted into the S region. This process explains the transmission of subgap-energy carriers across an NS interface and results in the conductance enhancement for $V < 2\Delta/e$, where the factor 2 is from the two NS interfaces existing in a SGS junction. The subgap structure with multiple $dI/dV$ peaks represents the multiple Andreev reflection (MAR) [14], in which the peaks occur at $V = 2\Delta/ne$ with an integer $n$ referring to the order of the AR. The $V$-axis position of the $n = 1$ MAR peak allows one to estimate the superconducting energy gap of Al electrodes to be $\Delta \sim 55$ μeV, which is about a factor of two smaller than other reports [5,15]. The reduced $\Delta$ is also manifested by the suppressed $T_c = 0.38$ K of Al, where the relation of two quantities are given by the Bardeen-Cooper-Schrieffer theory in the weak-coupling limit as $\Delta = 1.76\ k_B T_c$, where $k_B$ is the Boltzmann's constant [12, 16]. Other devices used in this study with thicker Al electrodes ($t_{Al}$>90 nm) recover the values of $T_c$ (~0.84 K) and $\Delta$ (~100 μeV) at the base temperature. Thus, the reduction of $T_c$ or $\Delta$ is attributed to the enhanced surface scattering in thin Al electrodes ($t_{Al}$ = 70 nm).

While the supercurrent and the subgap structure of $dI/dV$ are well understood by the conventional superconducting proximity effect and the phase-coherent Andreev reflection, there occurs "above-gap" structure of $dI/dV$ in a high voltage bias above the sum-gap voltage. Figure 3(a) shows multiple $dI/dV$–$V$ curves with different $V_g$ in a wide range of $V$. For $V \gg 2\Delta/e$, we observed a conductance jump accompanied by a jump in $dI/dV$, the $V$ position of which is highly sensitive to $V_g$. The above-gap structure is also evident in the $I$-$V$ curve as a cusp, as shown in Fig. 3(b). The characteristic voltage ($V^*$) and current ($I^*$) positions of these cusps, denoted by the arrows, vary with $V_g$. Here, we also note that the $V^*$ and $I^*$ for the above-gap structure are inversely proportional to each other. The $V_g$ dependence of $V^*$ (square symbols) and $I^*$ (circle symbols) are shown in Fig. 3(c), which reveals an opposite correlation between the two parameters. The behavior of $V^*$ and $I^*$ was similar in all devices used in this study. Figure 3(d) shows that the value of the characteristic power $P^*$ ($=I^*V^*$) turns out to be almost constant as $P^* = 0.47$ nW for the device D1 (refer to Table I for other devices), irrespective of the value of $V_g$. Similar behavior of the above-gap structure is observed in all our SGS devices, but with sample-dependent value of the characteristic power. For D2, for instance, with higher $T_c$ (= 0.82 K) of Al electrodes, $P^*$ becomes constant at much higher value of ~4.32 nW at $T$ = 10 mK and in zero magnetic field [see the arrow in the inset of Fig. 3(d)].

To find the physical cause of the above-gap anomaly, we investigated $T$ and $H$ dependences of the characteristic power $P^*$. In contrast to the insensitivity of $P^*$ to the variation of $V_g$, $P^*$ shown in Figs. 4(a) and (b) are highly sensitive to $T$ and $H$, respectively. The value of $P^*$ decreases with increasing $T$ and vanishes near $T_c$ of Al, showing a temperature dependence similar to that of $\Delta$ of the superconducting electrodes [see the arrow in Fig. 4(a)]. The inset of Fig. 3(d) shows similar $T$ dependence of $P^*$ of the device D2 for different backgate voltages. Fig. 4(b) shows a gradual decrease of $P^*$ with increasing $H$ at the base temperature of $T = 10$ mK and vanishes at $H_c$ [see the arrow in the Fig. 4(b)], which corresponds to the magnetic-field suppression of $T$ or $\Delta$ of the Al electrodes.

In connection with our results, a hysteretic behavior was reported recently in the $I$-$V$ characteristics of sub-micrometer-scale SNS (Al/Cu/Al) Josephson junctions and was interpreted in terms of the increase of $T_e$ of its normal-metallic region N as the junction switches to the resistive state [17]. According to the studies [17-20], electrons and phonons become thermally decoupled at low temperatures ($T < 1$ K), where the electron and phonon temperatures become considerably different. The electron temperature is predicted to follow the relation

$$T_e = (P/(\Sigma V_d) + T_{ph}^n)^{1/n}, \qquad (1)$$

where $P$ is the externally supplied power and $T_e$ ($T_{ph}$) is the mean electron (phonon) temperature in a sample with the normal-region volume $V_d$ of an SNS junction. Here, $\Sigma$ is a material-dependent parameter and $n$ is 5 for a three-dimensional free-electron-gas system [18] at low temperatures. In a two-dimensional system, the $V_d$ and $n$ are replaced by the area $A$ of the intermediate graphene region and 4 [21], respectively.

Since the above-gap anomaly takes place for a constant value of power $P^*$ ($= I^* V^*$) dissipated in the intermediate graphene region sandwiched between two S electrodes of an SGS junction, $P^*$ should be related to the enhanced electron temperature ($T_e$) in the region. As these hot electrons are injected to an Al superconducting electrode they thermalize a thin layer of Al at the interface, because electrons and phonons are no longer decoupled in Al electrodes. We assume that, for the characteristic power $P^*$, the electron temperature $T_e$ in the graphene region of an SGS junction increases up to $T_c$ of the superconducting electrodes. Then, the thermalized thin layer of Al at the GS interface loses the superconductivity, along with the disappearance of the AR at the interface and the consequent AR-enhanced conductance. In Fig. 5(a) the *I-V* curve of D1 for $V_g = 0$ is replotted. It shows the low-bias region of the enhanced conductance (Region I) by the AR process and the region of the complete suppression of the AR-induced conductance enhancement (Region III). Figure 5(b) clearly illustrates the *dI/dV* variation for the corresponding regions. Region II is the intermediate one between Regions I and III, where the corresponding *dI/dV* becomes even smaller than that of Region III as shown in Fig. 5(b). One notes that the jump in *dI/dV* takes place at the onset point of the voltage bias where the AR becomes completely suppressed.

To calculate the electron temperature $T_e$ in the graphene region of our SGS junction, we first find the value of the coefficient $\Sigma$ (~5.86 nW·μm$^{-2}$·K$^{-4}$) of the device D1 (refer to Table I for other devices) by using Eq. (1) by setting the values $T_e = T_c$ (= 0.38 K), $P = P^*$ (= 0.47 nW), $A$ = 1.3 x 3.2 μm$^2$, $T_{ph} \sim T_0$ (= 10 mK), where $T_0$ is the base temperature used for the $I_c$ measurements. Once the temperature-independent value of $\Sigma$ is obtained, we calculate the predicted temperature dependence of $P^*$ by using Eq. (1) for the base temperature varying from 10 mK to 0.40 K and compare it with that of observed $P^*$.

In Figure 4(a) the temperature dependence of observed $P^*$ (square symbols) is compared with that of $P^*$ predicted by Eq. (1) (the solid line) for the device D1. Values of the two sets of $P^*$ are rapidly reduced near $T_c$ of the Al electrodes as the superconductivity vanishes. The observed values of $P^*$ well follow the temperature dependence predicted by Eq. (1). As the base temperature $T_0$ is increased, less dissipation is required to raise $T_e$ up to $T_c$, which thus leads to a smaller value of $P^*$. Similar argument is valid for *H* dependence of $P^*$ shown in Fig. 4(b). The *T* and *H* dependences of $P^*$ indicate that $P^*$ is only related to $T_c$ or $\Delta$ of the superconducting electrodes.

Now let's focus on the detailed feature shown in Fig. 5. For $V < 2\Delta/e$, the MAR occurs at both interfaces between graphene and the superconducting electrodes as depicted in Fig. 2(d), which gives the abrupt conductance peaks at voltages that satisfy the MAR condition. For $V > 2\Delta/e$, the MAR condition is no longer satisfied. Even in this case, however, separate Andreev reflection occurs at each interface, giving rise to the sustained excess conductance. For an ideal SNS junction, with a full Josephson supercurrent ($I_c \sim \pi\Delta/2eR_N$; $\Delta$ is the superconducting gap and $R_N$ is the normal-junction resistance) corresponding to highly transparent interfaces, the excess conductance would be present in all the bias range for $V > 2\Delta/e$. When the junction supercurrent is much reduced than the full ideal value as observed in the device D1 [see Fig. 5(a)] the excess conductance in low bias is suppressed but gradually recovers the value corresponding to the Andreev-reflection-induced enhanced conductance at both interfaces [Region I of Fig. 5(a)]. In our device, entering into Region II, the excess conductance starts being reduced and completely disappears at $V^*$, the boundary bias value between Regions II and III. We believe that the reduction of the excess conductance in Region II is caused as the hot electrons entering an electrode thermalize the thin superconducting layer at the interface while suppressing the superconducting gap in the layer. The thermalization thus induces a decrease of the portion of carriers that are Andreev-reflected for $V<2\Delta/e$ at an interface. The excess conductance disappears completely at $V^*$, where the superconducting gap fully closes as $T_e$ reaches $T_c$. This picture explains the general feature of the differential conductance in Fig. 5(b). But, at the same time, we also notice in Figs. 5(a) and 5(b) that the reduction rate of the excess conductance remains finite as the bias approaches $V^*$, giving rise to the jump in the differential conductance. To understand this feature one needs to note that, as the bias increases, both the number and the energy (or $T_e$) of hot electrons increase, which more effectively suppresses $\Delta$ in the interfacial layer than the case of the simple increase of the number of hot electrons with a fixed $T_e$. This may induce a faster decrease of the excess conductance as the bias approaches $V^*$, resulting in the differential conductance below the normal-state junction value in Region III. Then, an abrupt increase of the differential conductance follows at $V^*$, as the superconducting gap fully closes. The abrupt change of the differential conductance points that the thermalization of the NS interface occurs at once all over the interface in the device D1. The differential conductance change turns out to be more gradual in device D2 (not shown), for instance, near the boundary between Regions II and III, where a spatial distribution in the bias value of $V^*$ is supposed to be present at the interface.

The insensitivity of $P^*$ to $V_g$ indicates the existence of a critical Joule power, which is sufficient to convert the thin layer of a superconducting electrode at a GS interface to normal. The value of $P^*$ is not dependent on the normal-conductance ($R_n^{-1}$) of the graphene layer but on the superconducting parameter, $T_c$ or $\Delta$, of superconducting electrodes. For instance, almost 7 times lager $P^*$ of D3 (3.53 nW) than that of D4 (0.51 nW) was observed in spite of very similar device geometry with each other (size of graphene layers are 1.3 x 1.0 $\mu m^2$ and 1.3 x 1.2 $\mu m^2$, respectively). This can be understood by much higher $T_c$ of D3 ($T_c \sim 0.84$ K) than that of D4 ($T_c \sim 0.38$ K), the difference of which was caused by the difference in the thickness of Al layer; 90 nm for D3 and 70 nm for D4.

Similar feature of the above-gap anomaly was observed in other conventional SNS proximity-coupled junctions consisting of a superconductor and a semiconductor quantum well [22] or a superconductor and a normal metal [23-24]. This invites more general interests in this anomalous phenomenon. The former observation was interpreted in terms of the multiple normal reflections in the semiconducting layer of the Andreev-reflected holes. This scenario, however, cannot be applied to our SGS junction geometry, which prohibits the multiple normal reflection of Andreev reflected holes inside mono-atomic layer of graphene under the Al electrode. The interpretation for the latter observation rests on a discontinuous change of the local contact resistance due to a large bias current [23]. In our experiments, however, the characteristic bias current for the anomaly $I^* = 1.8$ μA corresponds to a current density of $j^* = 5.14$ x $10^3$ A/cm$^2$ (with $t_{Al} = 0.07$ μm, $W_{Al} = 0.5$ μm for D1), which is three orders of magnitude smaller than the critical value used in Ref. 23. In this study, we were able to examine the variation of $I^*$ and $V^*$ at a given temperature by measurements at different backgate voltages, which enabled us to confirm that the above-gap anomaly corresponded to a constant power dissipation in the normal layer (graphene layer in this study). In this sense, our graphene-based proximity junction devices provide a very unique system to clarify the controversial origin of the above-gap anomaly, which is often observed in diverse SNS proximity junctions.

## IV. CONCLUSION

In summary, we have successfully fabricated the superconducting junctions of graphene and observed the supercurrent flow at low temperature below $T_c$. The electrical transport across a junction in low bias of $V < 2\Delta/e$, such as a modulation of $I_c$ as a function of $H$ or $V_g$, is well understood in terms of the superconducting proximity effect and Andreev reflection. In high bias of $V >> 2\Delta/e$, however, an anomalous jump in the junction differential conductance was observed. The $V_g$ dependence of $P^*$ indicates that the anomaly corresponds to constant dissipation $P^*$ in the graphene layer. The $T$ and $H$ dependences of $P^*$ indicate that the anomaly takes place as the electron temperature increases up to $T_c$ of the superconducting electrodes and suppresses their superconductivity, along with the reduction of the AR-induced conductance enhancement. Thus, this phenomenon is well understood by the self-heating and a consequent increase of the electron temperature $T_e$ in graphene. The result indicates that a serious consideration is required for the electron heating effect on the quantum electronic transport at very low temperature, in particular, for the graphene nanoribbon containing very small lateral area of the junction.

**ACKNOWLEDGMENTS**

One of us (JHC) wishes to acknowledge useful discussion with Dr. M.-H.Bae, S.-G. Nam, and Dr. J. S. Jung. This work was supported by National Research Foundation of Korea through Acceleration Research Grant (No. R17-2008-007-01001-0).

Table I. Device parameters. $W$ and $L$ are the width and length of the intermediate graphene layer in SNS devices. $T_c$ is the superconducting critical temperature of the Al electrodes. Ti/Al/Au indicates the thicknesses of the Ti, Al, and Au mutilayer electrode. $P^*$ is the characteristic dissipation where the above-gap anomaly takes place. $\Sigma$ is the coefficient in Eq. (1).

| Device | $W$ (μm) | $L$ (μm) | $T_c$ (K) | Ti/Al/Au (nm) | $P^*$ (nW) | $\Sigma$ (nW μm$^{-2}$ K$^{-4}$) |
|---|---|---|---|---|---|---|
| D1 | 3.2 | 0.30 | 0.38 | 10/70/5 | 0.47 | 5.86 |
| D2 | 7.4 | 0.45 | 0.82 | 10/90/5 | 4.32 | 0.36 |
| D3 | 1.0 | 0.30 | 0.84 | 10/90/5 | 3.53 | 5.41 |
| D4 | 1.2 | 0.30 | 0.38 | 10/70/5 | 0.51 | 15.75 |

**FIGURE CAPTIONS AND FIGURES**

**Figure 1.**

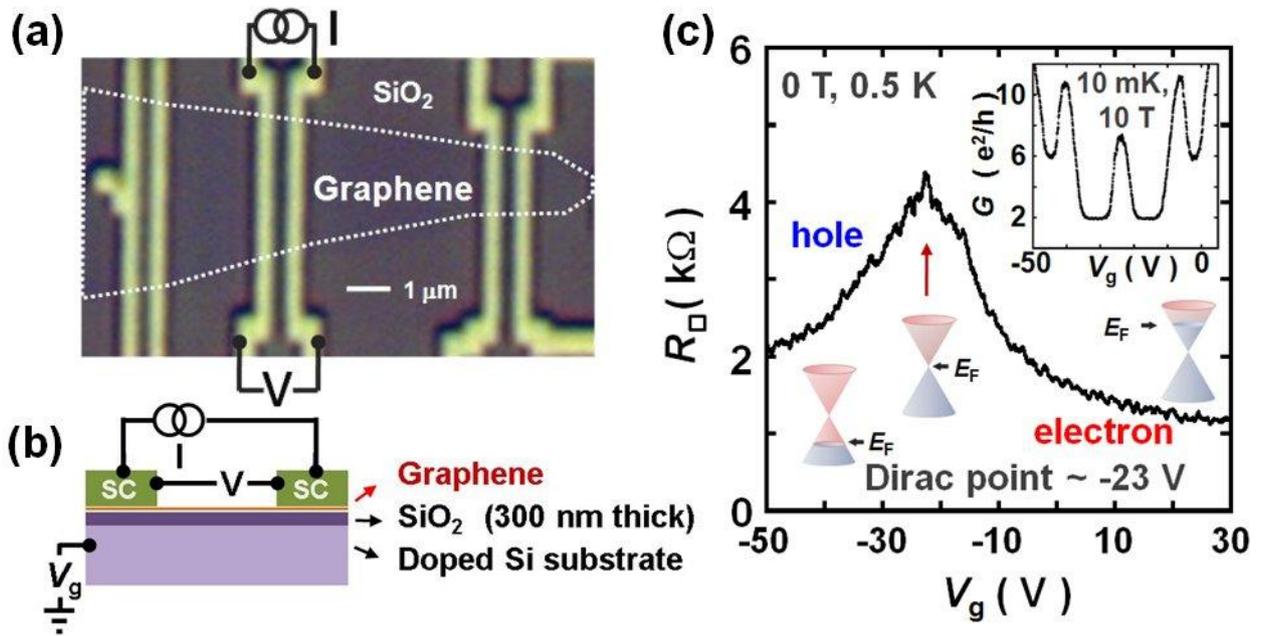

Figure 1. (a) An optical-microscopic image of our device D1. (b) Schematic cross-sectional view of the measurement configuration for our superconductor-graphene-superconductor (SGS) junction. (c) Variation of the two-terminal sheet resistance ($R_\square$) as a function of the backgate voltage ($V_g$) at 0.5 K.

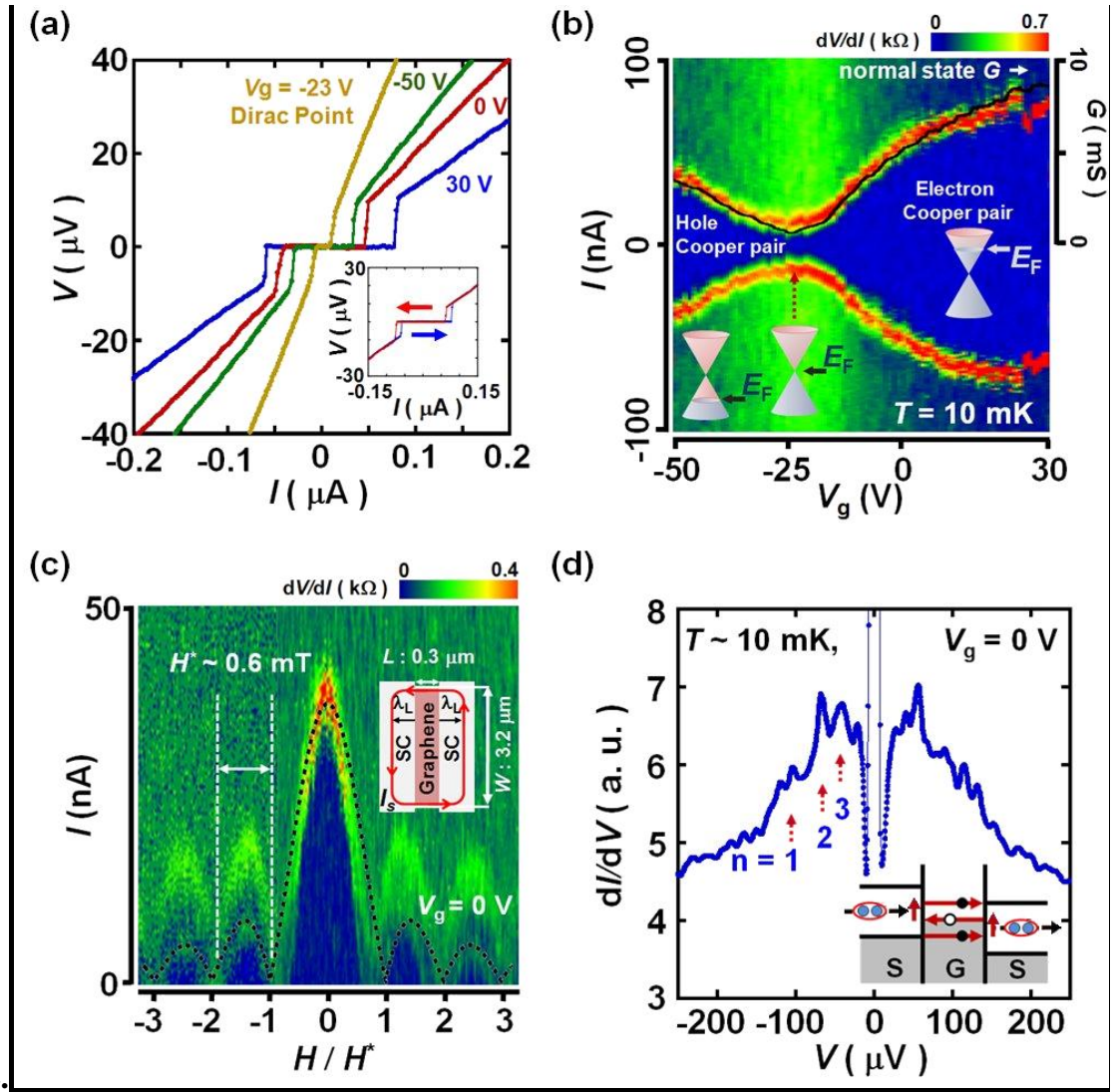

**Figure 2.**

Figure 2. (a) The current-voltage (*I-V*) characteristics at various backgate voltages $V_g$, showing a modulation of the critical current. Inset: *I-V* characteristics with the current bias swept up and down. The asymmetry of the data in the main panel is due to the hysteretic behavior. (b) Color-coded plot of *dV/dI* as a function of $V_g$ and *I*. The deep blue color stands for the zero-resistance supercurrent region. The current was swept from negative values [see also inset to Fig. 1(a)]. (c) Color-coded plot of the differential resistance (*dV/dI*) as a function of current bias (*I*) and magnetic field (*H*) at *T* = 10 mK (the deep blue color corresponds to zero-differential-resistance (*dV/dI*) Josephson supercurrent state of the SGS junction. (d) The differential conductance (*dI/dV*) vs sample voltage (*V*) showing the multiple-Andreev-reflection (MAR) peak. Inset: schematic illustration of the MAR process for *n*=3.

**Figure 3.**

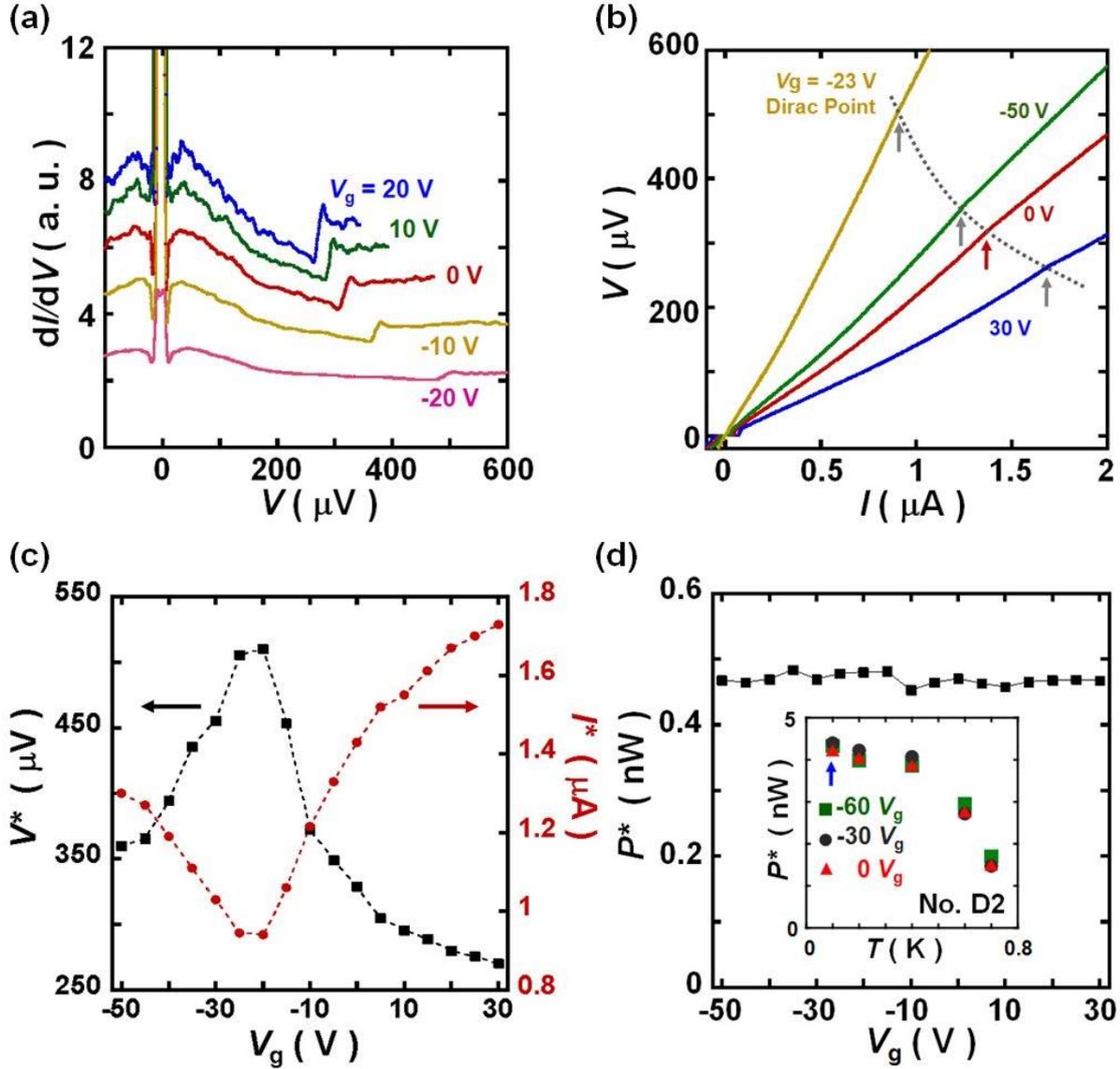

Figure 3. (a) $dI/dV$ vs $V$ curves in a wide range of $V$ for different $V_g$, showing gradual suppression of the Andreev reflection with $V$ below $V < 2\Delta/e$ and a conductance jumps at a high-bias voltage of $V = 2\Delta/e$. (b) $I$-$V$ characteristics for several different values of $V_g$ with a cusp in a high bias above the sum gap voltage $V = 2\Delta/e$, for all values of $V_g$. (c) The $V_g$ dependence of the characteristic voltage $V^*$ and current $I^*$ for the jump or the cusp. (d) The $V_g$ dependence of the characteristic power $P^*$ (= $I^*V^*$) at $T = 10$ mK. Inset: $P^*$ vs $T$ for different $V_g$, obtained from the device D2.

**Figure 4.**

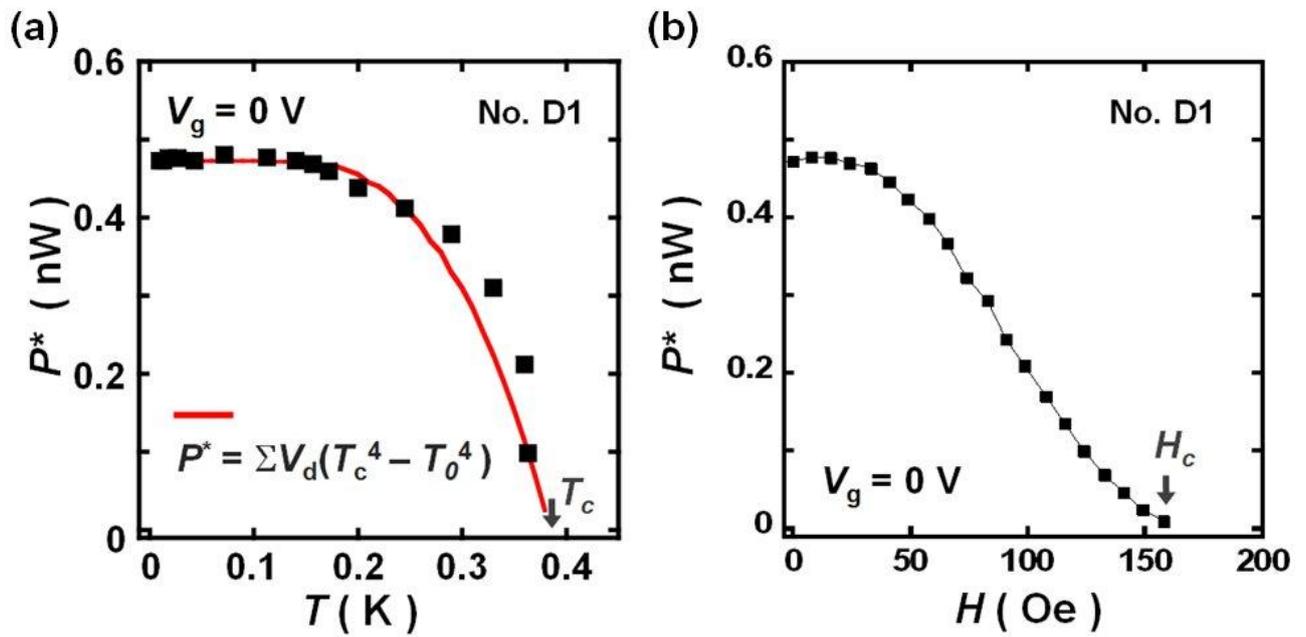

Figure 4. (a) Temperature dependence of $P^*$ (square) and the calculation result (solid line) based on Eq. (1), details of which are explained in the text. (b) Perpendicular-magnetic-field ($H$) dependence of $P^*$. The solid line is a guide to eyes.

**Figure 5**

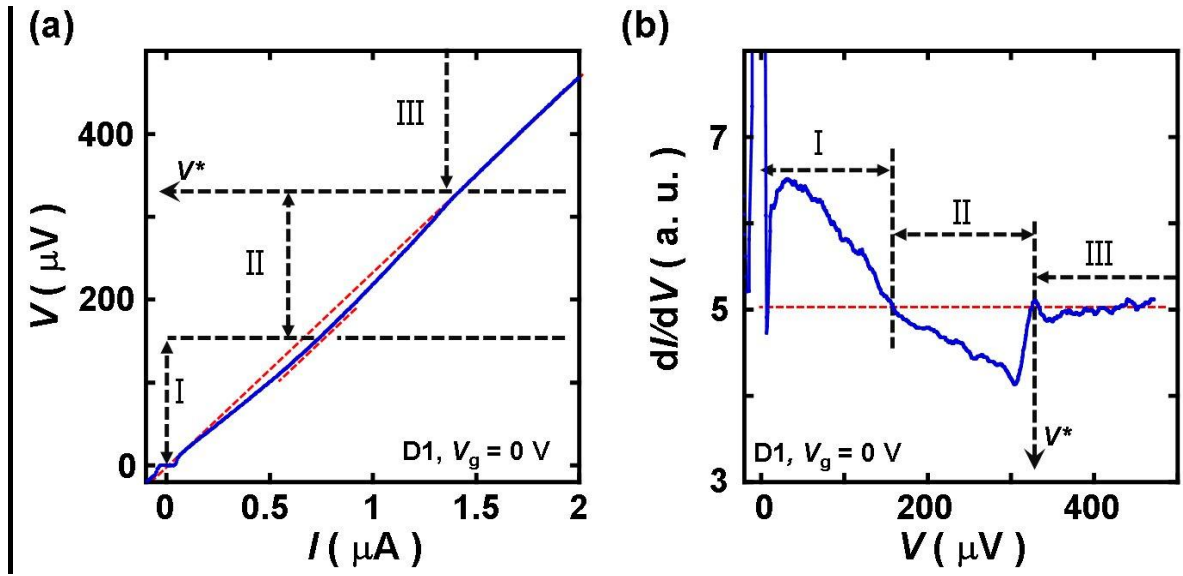

Figure 5. (a) A replot of the *I–V* characteristics of the device D1 in Fig. 3(b) for $V_g = 0$. Regions I, II, and III represent the regions of AR-induced enhanced conductance, intermediate state, and the complete suppression of conductance enhancement. The dotted line (gray line) represents the normal-junction *I-V* curve (*I-V* curve for an SNS junction with an ideally high junction critical current). (b) A replot of the corresponding differential conductance in Fig. 3(a) for $V_g = 0$. The above-gap anomaly takes place at the boundary of Regions II and III.